\documentclass[%
 reprint,
 amsmath,amssymb,
 aps,
]{revtex4-1}

\usepackage{bibentry}
\usepackage{graphicx}
\usepackage{dcolumn}
\usepackage{bm}
\usepackage[mathlines]{lineno}

\begin{document}

\preprint{APS/123-QED}

\title{Optical transport and manipulation of an ultracold atomic cloud using focus-tunable lenses}

\author{Julian L\'{e}onard}
\author{Moonjoo Lee}
\author{Andrea Morales}
\author{Thomas M. Karg}
\author{Tilman Esslinger}
\author{Tobias Donner}

\email{donner@phys.ethz.ch}
\affiliation{Department of Physics, ETH Z\"urich, 8093 Z\"urich, Switzerland}%

\date{\today}

\begin{abstract}
We present an optical setup with focus-tunable lenses to dynamically control the waist and focus position of a laser beam, in which we transport a trapped ultracold cloud of \textsuperscript{87}Rb over a distance of $28\,\mathrm{cm}$. The scheme allows us to shift the focus position at constant waist, providing uniform trapping conditions over the full transport length. The fraction of atoms that are transported over the entire distance comes near to unity, while the heating of the cloud is in the range of a few microkelvin. We characterize the position stability of the focus and show that residual drift rates in focus position can be compensated for by counteracting with the tunable lenses. Beyond being a compact and robust scheme to transport ultracold atoms, the reported control of laser beams makes dynamic tailoring of trapping potentials possible. As an example, we steer the size of the atomic cloud by changing the waist size of the dipole beam.
\end{abstract}

\maketitle

The control of cold atomic gases with dipole potentials produced by far-off-resonance laser beams has proven to be a uniquely powerful tool in the production and manipulation of quantum gases \cite{grimm2000}. Recently, substantial progress has been made by introducing experimental techniques like high-resolution microscopy \cite{bakr2010, sherson2010} or holographic beam-shaping \cite{bakr2009} to create atomic clouds in locally structured \cite{ramanathan2011, brantut2012} or box-like potentials \cite{gaunt2013}. Yet, the ability to change optical potentials dynamically is not very advanced. However, the most widely used implementation of dynamical potentials is the transport of ultracold atoms between two spatially separated regions of an experiment, with one region optimized for the production of the cold atomic cloud and the other region optimized to investigate the cloud \cite{gustavson2001}. With the increasing interest in hybrid setups coupling cold gases to other quantum systems \cite{brennecke2007, camerer2011, vetsch2010}, the transport of atomic clouds now has become a crucial technique.

Optical transport can be performed by displacing the focus of a dipole beam in which the atoms are trapped. This has been achieved with the focussing lens mounted on an air-bearing translation stage \cite{gustavson2001}. Compared to magnetic transport, where the trap center is shifted either by a chain of overlapping coil pairs through which a current is applied sucessively \cite{greiner2001}, or by physically moving one coil pair \cite{goldwin2004}, this avoids to surround the vacuum chamber with a number of magnetic coils limiting the optical access. Yet, it comes with the drawback of placing an expensive and cumbersome translation stage close to the vacuum chamber, which bears the risk of transferring vibrations to the dipole trap or the optical table. Another method relies on trapping the atomic cloud in a one-dimensional optical lattice, which is then turned into a moving standing wave by detuning the frequencies of two counterpropagating beams with respect to each other \cite {sauer2004}. However, for Gaussian beams this technique is only applicable on short distances or in the vertical direction, because of the weak radial confinement compared to gravity \cite{schmid2006}.

\begin{figure}[b]
	\centering
		\includegraphics[width=\columnwidth]{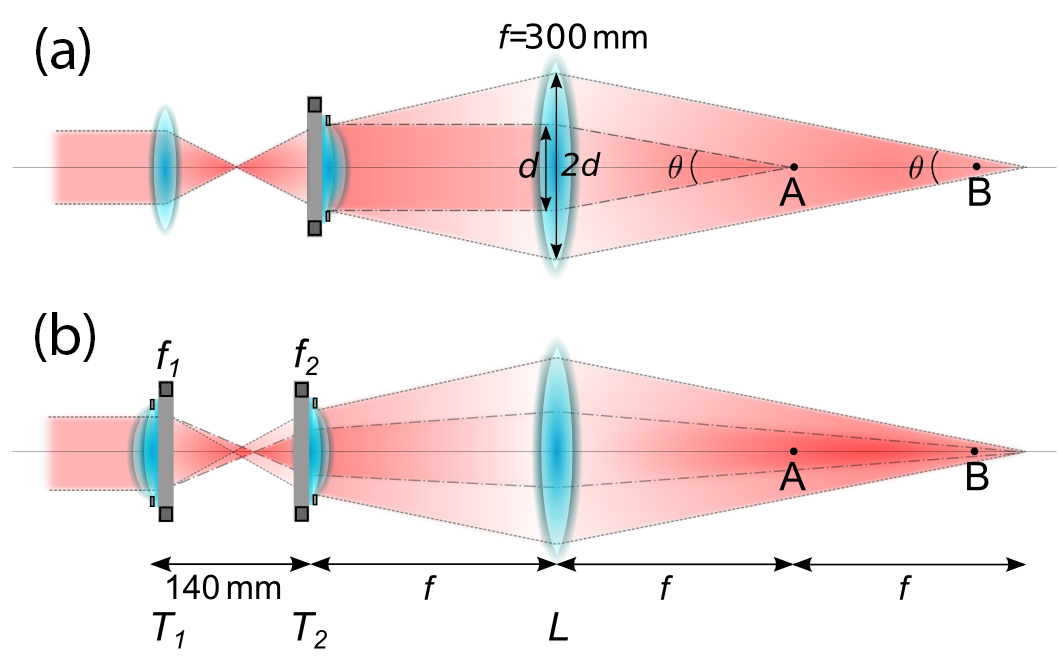}
	\caption{Setup to dynamically control size and position of a dipole trap. (a) Transport at constant waist over a distance of $28\,\mathrm{cm}$. If the separation between the tunable lens $T_2$ and the static lens $L$ equals the focal length $f$ of the latter, the two beams can be transformed into each other by tuning $f_2$, while maintaining the same divergence $\theta=d/f$, thus the same waist size, between A and B. (b) Independent control over waist size and position of the focus. Replacing the first lens with a tunable lens $T_1$ allows to change the beam size at $T_2$, resulting in a different divergence behind $L$. }
	\label{fig:setup}
\end{figure}

\begin{figure*}[t]
	\centering
		\includegraphics[width=0.79\textwidth]{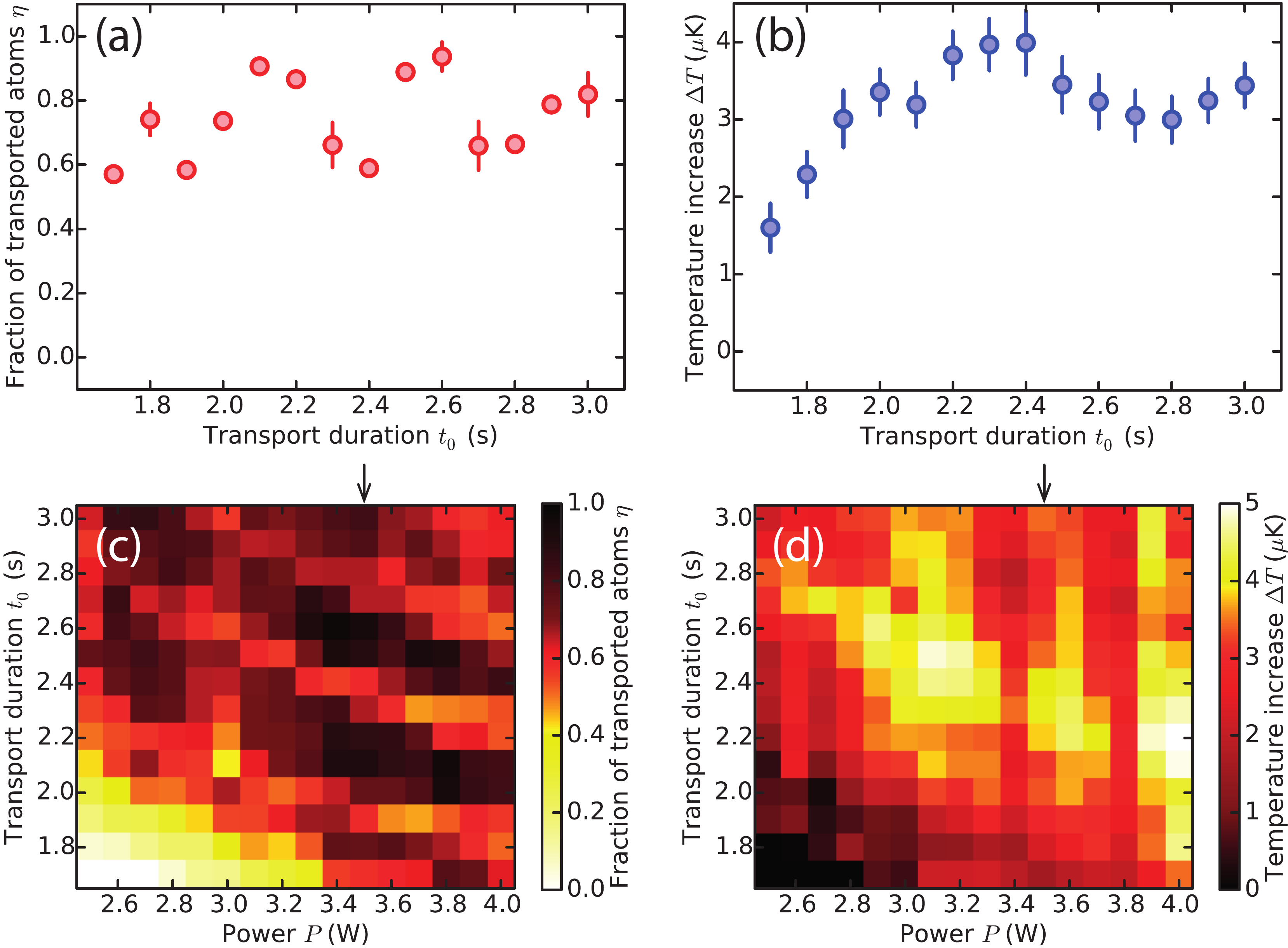}
	\caption{Characterization of the optical transport.  (a) The transfer fraction $\eta=N_B/N_A$ and (b) the increase in temperature $\Delta T=T_B-T_A$ are shown for a dipole beam power of $P=3.5\,\mathrm{W}$. Each data point represents the average of three measurements. The errorbars denote the statistical error including the standard deviations of the measurements before and after the transport and, for (b), the uncertainty of the heating rate. Panels (c) and (d) show the full parameter scan in transport duration $t_0$ and dipole beam power $P$. The small arrows indicate the columns for the line cuts in (a) and (b).}
	\label{fig:transport}
\end{figure*}

Here, we present a different approach based on focus-tunable lenses, which allows to move the position of a dipole trap and additionally to tune its size. It combines a compact arrangement with a static setup. The rapid progress in the development of tunable and adaptive optics now allows to fabricate high-grade focus-tunable lenses, establishing them in an increasing number of fields of research and industry, such as material processing \cite{Eberle2013}, photography \cite{Savidis2013}, trapped nanoparticles \cite{Tanaka2013} or bioimaging \cite{grewe2011}. We use tunable lenses of the type EL-10-30 from the supplier \textit{Optotune}. The lens surface is spherical with an aperture of $10\,\mathrm{mm}$ and a wavefront error of 0.1--0.2$\,\lambda$ (depending on the focal length), where $\lambda=1064\,\mathrm{nm}$ is the anti-reflection coating wavelength. The body of the lens is filled with a low-optical absorption liquid and the surface is sealed off with an elastic polymer membrane. As indicated in the sketches in Fig.\,\ref{fig:setup}, an applied current in a coil can increase the membrane curvature by a magnetic ring mechanically pressing liquid from the outer area to the lens center. Thus the focal length can be tuned within a range of 40--140$\,\mathrm{mm}$ \cite{datasheet}.

We use these lenses to transport a cloud of ultracold atoms over a distance of $28\,\mathrm{cm}$ from a first vacuum chamber A, where the atomic cloud is prepared, to a second one B, which is our science chamber, as shown in Fig.\,\ref{fig:setup}. In principle, a focus displacement can be achieved using a single tunable lens focussing a collimated beam. However, increasing the focal length increases the waist size as well, thereby changing trapping frequencies and trap depth during the transport. Instead, the setup presented here provides uniform trapping conditions over the full transport distance. This is preferrable, since only high confinement and large trap depth allow for fast transport. The two beams in Fig.\,\ref{fig:setup}\,(a) are focussed behind the static lens $L$ with focal length $f=300\,\mathrm{mm}$ at distances $f$ and $2f$. Their waist sizes are equal if their divergences are. This requires beam diameters of $d$ and $2d$ at $L$, respectively, resulting in the same divergence of $\theta = d/f$ for both. The beam with the focus at $f$ must be collimated before the lens, and the beam with the focus at $2f$ must have the same divergence $\theta$ before passing $L$. Therefore the two beams have the same size at a distance $f$ before $L$. Placing a lens $T_{\mathrm{2}}$ with tunable focus $f_2$ at this position allows to continuously transform one beam into the other, resulting in a moving focus at constant waist. Since $f_2>0$ the divergence can only be reduced and the beam must be focussed before crossing $T_2$. This can be achieved by a first static lens, which in turn defines the waist size behind $L$. In order to even gain independent control over position and waist size of the focus, the first static lens can be replaced by a tunable lens $T_1$, as shown in the extended setup in Fig.\,\ref{fig:setup}\,(b). Calculation of the Gaussian beam propagation through the system enables us to compute waist size and focus position for any focal length tuple $(f_1, f_2)$.

This extended setup is used in the following, choosing the distances as indicated in Fig.\,\ref{fig:setup}. We send a collimated laser beam with a $1/e^{2}$ diameter of $5.9\,\mathrm{mm}$ and a wavelength of $\lambda = 1064\,\mathrm{nm}$ out of a photonic crystal fiber to the tunable-lens setup. The lenses are steered via low-noise current sources. The focal lengths are first set to $f_1^{A}=52\,\mathrm{mm}$ and $f_2^{A}=88\,\mathrm{mm}$, resulting in a measured dipole beam waist size of $w_0=47.0\,\mathrm{\mu m}$ with the focus at the position A. For a dipole beam power of $P=4.0\,\mathrm{W}$, the expected radial trapping frequency and trap depth are $\omega_r/2\pi=802\,\mathrm{Hz}$ and $U_0=k_B \times 147\,\mathrm{\mu K}$, respectively. 

The ultracold atomic cloud is generated from initially $5\times 10^{9}$ \textsuperscript{87}Rb atoms in a three-dimensional magneto-optical trap (MOT) loaded from a two-dimensional MOT. After evaporative cooling in a hybrid trap formed out of the dipole beam potential and a magnetic quadrupole trap \cite{Lin2009}, we entirely switch off the magnetic potential and end up with $N_A=1.86(6)\times 10^{6}$ atoms at a temperature of $T_A=4.9(3)\,\mathrm{\mu K}$ in the dipole trap. Choosing a dipole beam power of $P=4.0\,\mathrm{W}$, we measure the longitudinal trapping frequency by modulating the trap position. Different from parametric heating, position-induced heating appears at the trapping frequency itself \cite{gehm1998}. We apply a sinusoidal modulation of $f_2^{A}$ with an amplitude of $\Delta f_2^{A}=0.4\,\mathrm{mm}$, resulting in a real-space amplitude of $\Delta z=4.5\,\mathrm{mm}$. Using time-of-flight absorption imaging along the dipole beam axis, we observe an increase in temperature around the trapping frequency $\omega_{z}/2\pi=1.72(2)\,\mathrm{Hz}$. Compared to established methods to measure trapping frequencies, such as parametric heating \cite{friebel1998} or excitation of the dipole mode \cite{Brzozowski2010}, this approach gives a clearer signal. The fast response time of the lens of a few milliseconds \cite{datasheet} should allow our technique to be applicable to trapping frequencies up to several hundred hertz.

In order to check for heating of the atomic cloud due to vibrations of the tunable lens setup at a static position, we hold the cloud in the dipole trap for up to $10\,\mathrm{s}$ and measure its temperature for different laser powers of 2.5--4.0$\,\mathrm{W}$. We observe heating rates of 60--240$\,\mathrm{nK/s}$, overall increasing with laser power, which lie about a factor of two to six above the expected spontaneous emission rates. In light of our low longitudinal trapping frequency $\omega_{z}$, we attribute this additional heating to the susceptibility of the system to residual noise in the hertz range.

Tuning the focal length of $T_2$, we now move the cloud position. While keeping $f_1=52\,\mathrm{mm}$ constant, we apply an s-shaped position profile (parabolic velocity profile) from $f_2^{A}=88\,\mathrm{mm}$ to $f_2^{B}=138\,\mathrm{mm}$ and subsequently measure the temperature at position B by time-of-flight absorption imaging perpendicular to the transport axis. We confirm the temperature and atom number measurements in both imaging systems to be compatible within each other by imaging the cloud at A, after the transport at B and after a double transport back at A. We do not observe any systematic difference in atom number and temperature within our reproducibility of $4\,\%$ and $6\,\%$, respectively. The imaging is carried out directly after the transport is completed, neglecting residual dipole oscillations that damp out with a time constant of $\tau_{\mathrm{dipole}}=0.45(6)\,\mathrm{s}$. 

We start our transport measurements at a fixed dipole beam power of $P=3.5\,\mathrm{W}$. In Fig.\,\ref{fig:transport}\,(a), we show the transfer fraction $\eta = N_B/N_A$ as a function of the transport duration $t_0$. Depending on $t_0$, transfer fractions close to unity can be reproducibly obtained. Fig.\,\ref{fig:transport}\,(b) shows the increase in temperature due to the transport, $\Delta T = T_B-T_A$, where in addition the previously calibrated heating rate for a static trap has been subtracted to get access to the pure heating caused by transporting the atoms. For all values of $t_0$, the heating lies in the range of a few mikrokelvin, typically around $3\,\mathrm{\mu K}$. We emphasize that since this data represent a difference, the actual standard deviation of the temperature data values is smaller than the errorbars apparently suggest. For both atom number and temperature we observe the same reproducibility as before the transport, implying that the tunable-lens setup does not alter the reproducibility of our experimental conditions.

\begin{figure}[t]
	\centering
		\includegraphics[width=0.84\columnwidth]{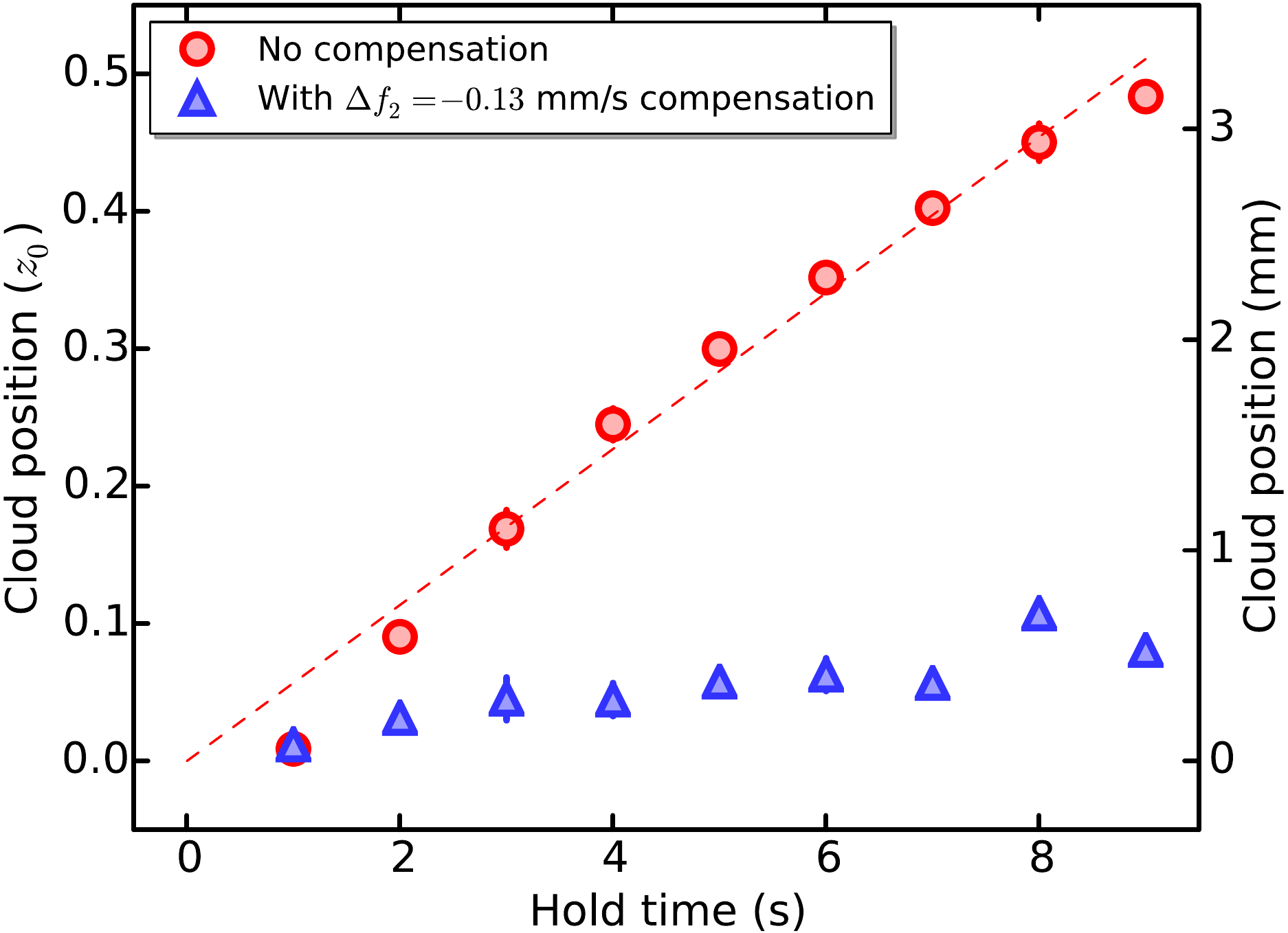}
	\caption{Longitudinal drift of the cloud center position. The red circles and the blue triangles represent the cloud position without and with the compensated drift by tuning the focal length $f_2$ over time. The dashed red line is a linear fit to the cloud position, giving a drift rate of $5.7(3)\,\% \times z_{0}/\mathrm{s}$, where $z_0=6.5\,\mathrm{mm}$ denotes the Rayleigh length of the dipole beam. Each data point represents an average of five measurements, errorbars denote the statistical error of one standard deviation.}
	\label{fig:drift}
\end{figure}

To further characterize the transport behaviour and to prove the flexibility on the choice of parameters, we vary both the transport duration $t_0$ and the dipole beam power $P$ over a broad range. Fig.\,\ref{fig:transport}\,(c) shows the transfer fraction $\eta$ as a function of these parameters. We observe values close to unity, i.e. $\eta_{\mathrm{max}} = 97(3)\,\%$, when choosing sufficiently high laser powers and transport durations. For short transport durations and low laser powers, we observe almost no transported atoms. This can be explained by the small trap depth at low dipole beam powers, which is decreased below $k_BT_A$ if the acceleration during the transport gets high for short transport durations. The temperature increase $\Delta T$ is shown in Fig.\,\ref{fig:transport}\,(d). For large trap depths, we observe an increase in temperature of $\Delta T = $ 1--5$\,\mathrm{\mu K}$. For short transport durations and small laser powers, the finite trap depth comes again into play and acts as an evaporative cooling step for the atoms.

The plots in Figs.\,\ref{fig:transport}\,(c) and (d) both show diagonal patterns, indicating a highly nonlinear behaviour of $\Delta T$ and $\eta$ in the considered parameter space of $t_0$ and $P$. We attribute this to the nonadiabatic nature of our transport: Since the transport duration is comparable to the inverse of the trapping frequency, $t_0\sim 1/\omega_z$, the cloud does not follow adiabatically the trap position, but dipole oscillations are excited. Depending on the specific relation of $\omega_z$ and the transport duration, these dipole oscillations can be of different amplitude or even fully suppressed \cite{couvert2008}. They are converted to heat via collisions during the transport and still continue afterwards. In combination with the finite trap depth, the oscillations in return lead to atom loss.

\begin{figure}[t!]
	\centering
		\includegraphics[width=0.78\columnwidth]{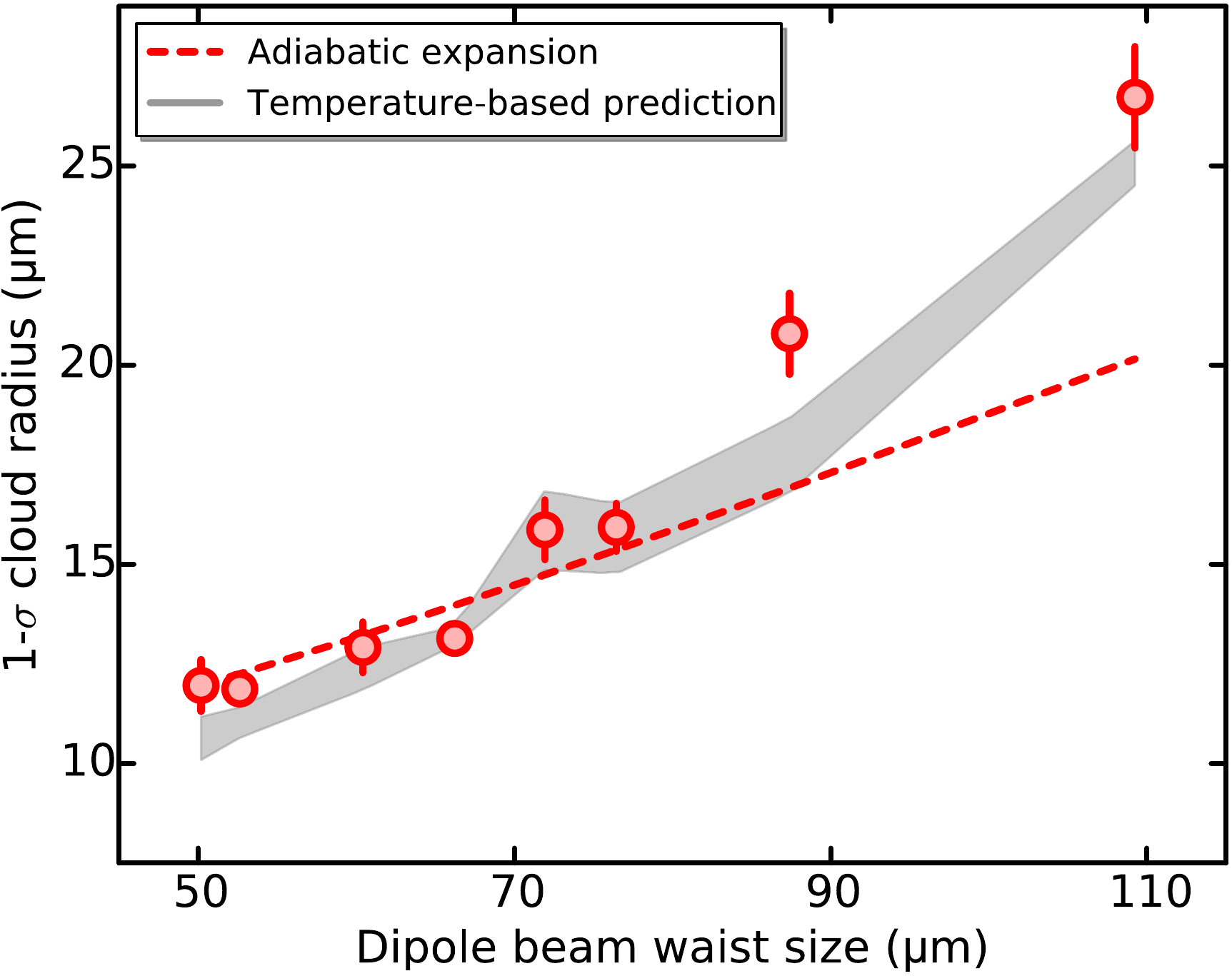}
	\caption{Dynamic control over the trapping potential. The cloud radius can be changed by increasing the waist size of the dipole beam trap at constant optical power (red data points). The grey shaded area spans the 2-$\sigma$ confidence interval of the expected cloud size based on the independently measured temperature, taking into account the finite resolution of our imaging setup with a numerical aperture of $\mathrm{NA}=0.06$. The red dashed line predicts the cloud radius assuming an adiabatic expansion of the trap. Each data point represents the average of five pictures, the errorbars denote the statistical error of one standard deviation.}
	\label{fig:size}
\end{figure}

The position stability of the dipole trap can be tested by measuring the drift behaviour for long hold times. After optical transport with the parameters $t_0=2.7\,\mathrm{s}$ and $P =3.8\,\mathrm{W}$, we hold the cloud for different durations and measure its position along the beam axis. The pictures are taken after up to $9\,\mathrm{s}$ hold times, starting at $1\,\mathrm{s}$ such that residual dipole oscillations are already damped out. As shown in Fig.\,\ref{fig:drift}, we observe a longitudinal drift that can be attributed to heating of the lens due to dissipated electric power. This changes the lens radius of curvature and thereby moves the focus position. The drift is expected from the data sheet \cite{datasheet}, but is reproducible under experimental conditions. Fitting the data with a linear function gives a rate of $0.37(2)\,\mathrm{mm/s}$, corresponding to $5.7(3)\,\%$ of the Rayleigh length $z_{0}=6.5\,\mathrm{mm}$ per second. As the drift rate is induced by the driving current of the lens, it depends on the focal lengths of the tunable lenses. We demonstrate that this drift can be overcome by constantly tuning the focal length $f_2^{B}$ in the opposite direction. During the hold time, we tune $f_2^{B}$ with the rate $\Delta f_2^{B}=-0.13\,\mathrm{mm/s}$, corresponding to a real-space change in focus position of $-0.37\,\mathrm{mm/s}$. As shown in Fig.\,\ref{fig:drift}, the drift rate can be almost entirely compensated while maintaining the reproducibility in position \footnote{New developments include an integrated temperature sensor allowing for active feedback on the focal length.}. We also check for position drifts in the radial direction when holding the cloud at a constant position, but observe no measurable drift rate within an uncertainty of $1\,\mathrm{\mu m}$. 

Our optical setup allows us to dynamically and independently change the trap curvature and depth, without affecting its position. So far, this was not possible, since  lowering the trapping frequency by decreasing the optical power via $\omega_{r, z} \sim \sqrt{P}$ was always accompanied by a smaller trap depth. Here, we tune the size of the cloud by changing the waist size with the focus-tunable lenses. During optical transport with $P=3.8\,\mathrm{W}$ and $t_0=2.7\,\mathrm{s}$, we apply s-shaped ramps on both focal lengths such that we stop at different tuples $(f_1^B, f_2^B)$, corresponding to different waist sizes of 50--110$\,\mathrm{\mu m}$ at position B. The radial cloud size is subsequently measured by \textit{in situ} absorption imaging. As shown in Fig.\,\ref{fig:size}, we observe a significant enlargement of the cloud radius for increasing waist size. 

At constant temperature, the cloud radius depends quadratically on the waist size,  $\sigma_r \sim w_0^{2}$. Taking into account the adiabatic expansion of the cloud, the decreased temperature leads to a linear dependency \cite{ligare2010}. We observe good agreement of the measurements with that assumption for moderate waist sizes. Above waist sizes of $\sim 80\,\mathrm{\mu m}$, the cloud radius increases faster than linearly, hinting towards nonadiabatic expansion along the long axis of the trap. Including time-of-flight temperature measurements to deduce the actual extension of the cloud, we find good agreement of the cloud radii over the whole range of examined waist sizes. 


In conclusion, we have demonstrated optical transport of ultracold atoms using focus-tunable lenses, achieving transfer fractions close to unity at temperatures of a few microkelvin. Furthermore, our setup offers the unique possibility to tune the waist size of the dipole beam, which allows to dynamically change the radius of the cloud at constant trap depth. As a consequence, a large range of trap curvatures and densities can be explored to optimize evaporative cooling \cite{Kinoshita2005} and trap transfer. The latter is particularly interesting for loading optical lattices to avoid heating due to density redistribution \cite{greif2013, hart2014}. Another possible application would be to produce an interference pattern with tunable lattice constant \cite{williams2008} by injecting two off-centered beams into a tunable-lens setup. Finally, the unprecedented dynamic control over size, density and collision rate realizable with tunable lenses may be used to implement novel schemes for the production and manipulation of ultracold gases.

We would like to thank Laura Corman, Christian Zosel and Samuel H\"ausler for their contributions at an early stage of the experiment. This work is supported by NCCR-QSIT, TherMiQ and SQMS (ERC advanced grant).

\bibliography{bib}

\end{document}